\pdfoutput=1
\documentclass[12pt, epsfig]{article}
\usepackage{amsmath,amsfonts,amssymb,amsthm,amstext,amscd,eucal,graphicx,xcolor,endnotes}
\usepackage[all]{xy}
\usepackage{dsfont,epiolmec}
\usepackage{amsmath}
\usepackage{slashed,ccaption,cite}
\usepackage{epsfig,graphicx,tikz, mathrsfs, calligra}
\usepackage{ulem}
\usepackage[unicode=true,
 bookmarks=false,
 breaklinks=false,pdfborder={0 0 1},backref=false,colorlinks=true]
 {hyperref}
\hypersetup{
 pdfauthor={Daniel Grumiller, Shahin Sheikh-Jabbari },
 linkcolor=blue,urlcolor=magenta,filecolor=green,citecolor=red,pdfstartview=FitV,bookmarksopen=true}
\usetikzlibrary{arrows}
\usetikzlibrary{decorations}\usetikzlibrary{decorations.shapes}%\usetikzlibrary{decorations.snake}
\usetikzlibrary{decorations.pathmorphing,backgrounds,calc,shapes,arrows,shadows, snakes}
\tikzset{zigzag/.style={decorate,decoration=zigzag}}
\makeatletter \@addtoreset{equation}{section}

\makeatletter\renewcommand\section{\@startsection {section}{1}{\z@}%
                                   {-3.5ex \@plus -1ex \@minus -.2ex}%nn
                                   {2.3ex \@plus.2ex}%
                                   {\normalfont\large\bfseries}}
\renewcommand\subsection{\@startsection{subsection}{2}{\z@}%
                                     {-3.25ex\@plus -1ex \@minus -.2ex}%
                                     {1.5ex \@plus .2ex}%
                                     {\normalfont\bfseries}}

\renewcommand{\baselinestretch}{1.2}

\makeatletter
\DeclareFontFamily{OMX}{MnSymbolE}{}
\DeclareSymbolFont{MnLargeSymbols}{OMX}{MnSymbolE}{m}{n}
\SetSymbolFont{MnLargeSymbols}{bold}{OMX}{MnSymbolE}{b}{n}
\DeclareFontShape{OMX}{MnSymbolE}{m}{n}{
    <-6>  MnSymbolE5
   <6-7>  MnSymbolE6
   <7-8>  MnSymbolE7
   <8-9>  MnSymbolE8
   <9-10> MnSymbolE9
  <10-12> MnSymbolE10
  <12->   MnSymbolE12
}{}
\DeclareFontShape{OMX}{MnSymbolE}{b}{n}{
    <-6>  MnSymbolE-Bold5
   <6-7>  MnSymbolE-Bold6
   <7-8>  MnSymbolE-Bold7
   <8-9>  MnSymbolE-Bold8
   <9-10> MnSymbolE-Bold9
  <10-12> MnSymbolE-Bold10
  <12->   MnSymbolE-Bold12
}{}

\let\llangle\@undefined
\let\rrangle\@undefined
\DeclareMathDelimiter{\llangle}{\mathopen}%
                     {MnLargeSymbols}{'164}{MnLargeSymbols}{'164}
\DeclareMathDelimiter{\rrangle}{\mathclose}%
                     {MnLargeSymbols}{'171}{MnLargeSymbols}{'171}
\makeatletter
\newcommand{\be}{\begin{equation}}
\newcommand{\ee}{\end{equation}}
\newcommand{\bea}{\begin{eqnarray}}
\newcommand{\eea}{\end{eqnarray}}
\newcommand{\bse}{\begin{subequations}}
\newcommand{\ese}{\end{subequations}}
\newcommand{\beqa}{\begin{eqnarray}}
\newcommand{\eeqa}{\end{eqnarray}}
\newcommand{\beqar}{\begin{eqnarray*}}
\newcommand{\eeqar}{\end{eqnarray*}}
\newcommand{\bi}{\begin{itemize}}
\newcommand{\ei}{\end{itemize}}
\newcommand{\bn}{\begin{enumerate}}
\newcommand{\en}{\end{enumerate}}
\newcommand{\ba}{\begin{array}}
\newcommand{\ea}{\end{array}}
\newcommand{\bc}{\begin{center}}
\newcommand{\ec}{\end{center}}

\definecolor{darkgreen}{rgb}{0,0.3,0}
\definecolor{darkblue}{rgb}{0,0,0.3}
\definecolor{darkred}{rgb}{0.7,0,0}

\newcommand{\old}[1]{}%{\sout{#1}}

\begin{document}
\renewcommand{\baselinestretch}{1.2}  %Line spacing

\newcommand\cnote[1]{\textcolor{red}{\bf [C:\,#1]}}
\newcommand\dnote[1]{\textcolor{magenta}{\bf [D:\,#1]}}
\newcommand\snote[1]{\textcolor{blue}{\bf [S:\,#1]}}

\begin{titlepage}

\begin{flushright}%\vspace{-3cm}
{
%IPM/P-22/xxx\\
\today }\end{flushright}

\vspace*{0truecm}

\newcommand{\mytitle}{On Symplectic Form for Null Boundary Phase Space}

\begin{center}
\Large{\bf{\mytitle}}

\vspace*{5mm}

\Large{\bf{M.M.~Sheikh-Jabbari}}
\\

\normalsize
\bigskip

%{$^a$ \it Institute for Theoretical Physics, TU Wien\\ Wiedner Hauptstr.~8, A-1040 Vienna, Austria}
%\\
{%$^b$ 
\it School of Physics, Institute for Research in Fundamental
Sciences (IPM)\\ P.O.Box 19395-5531, Tehran, Iran}\\ e-mail:~\href{jabbari@theory.ipm.ac.ir}{jabbari@theory.ipm.ac.ir}

\end{center}
\setcounter{footnote}{0}

\vskip 1cm

%\bigskip

% ABSTRACT WORD LIMIT: 125 WORDS; currently: 108 words

\begin{abstract}
To formulate  gravity in spacetimes bounded by a null boundary,  an arbitrary hypothetical null surface,  boundary degrees of freedom (d.o.f) should be added to account for the d.o.f and dynamics in the spacetime regions excised behind the null boundary. In the $D$ dimensional example, boundary d.o.f are labelled by $D$ charges defined at $D-2$ dimensional spacelike slices at the null boundary. While boundary modes can have their own boundary dynamics, their interaction with the bulk modes is governed by flux-balance equations which may be interpreted as a diffusion equation describing ``dissolution'' of bulk gravitons into the boundary. From boundary viewpoint, boundary d.o.f obey local thermodynamical equations at the boundary. Our description suggests a new ``semiclassical'' quantization of the system in which boundary d.o.f are quantized while bulk is classical. This semiclassical treatment may be relevant to questions in black hole physics. 
\end{abstract}

%\dnote{word limit abstract: 125 words}

\vskip 3mm

\begin{center}{\textit{Prepared   in memory of Prof. T. Padmanabhan,\\ to appear in the Topical Collection (TC) of General Relativity and Gravitation (GERG).}}\end{center}

\end{titlepage}

{\large{\calligra{W}}}e typically face formulating physics problems in some specified regions of spacetime. The boundary which is a codimension one surface in $D$ dimensional spacetime may have null, timelike or spacelike sections. Boundaries may be hypothetical regions in spacetime or  physical surfaces; they may be at asymptotic regions of spacetime where spacetime is naturally limited to one side of the boundary or may be hypersurfaces dividing the spacetime into ``inside and outside'' or ``front and behind'' regions. In the latter case one may excise the region behind the boundary and try to formulate the problem in this excised spacetime. In this note we describe physics from the viewpoint of the ``front observer'' who does not have access to the behind region. This is essentially an update on ``Horizon 2020'' essay \cite{Grumiller:2020vvv}, which itself was a continuation of \cite{Sheikh-Jabbari:2016lzm}.

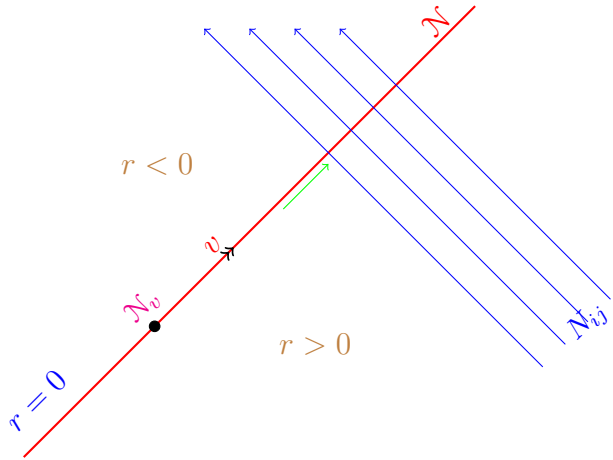
\begin{figure}[tp]
\def \L {3.0}
    \centering
% FIG starts below    
\begin{tikzpicture}
  % event horizon
  \draw[thick,red] (-\L,-\L) coordinate (b) -- (\L,\L) coordinate (t);
%\draw[blue,->] (1.8*\L,-0.1*\L) -- (0.8*\L,0.9*\L);          
  %\draw[blue,->] (1.7*\L,-0.2*\L) -- (0.6*\L,0.9*\L);           
  \draw[blue,->] (1.6*\L,-0.3*\L) -- (0.4*\L,0.9*\L);           
  \draw[blue,->] (1.5*\L,-0.4*\L) -- (0.2*\L,0.9*\L);
  \draw[blue,->] (1.3*\L,-0.6*\L) -- (-0.2*\L,0.9*\L);
  \draw[blue,->] (1.4*\L,-0.5*\L) -- (-0.0*\L,0.9*\L);
  \draw[black,thick,->] (-0.1*\L,-0.1*\L)--(-0.09*\L,-0.09*\L); \draw[black,thick,->] (-0.08*\L,-0.08*\L)--(-0.07*\L,-0.07*\L);
      \draw[red] (-0.1*\L,-0.0*\L) node[left, rotate=45] (scrip) {{$v$}};
%  \draw[blue] (1.1*\L,-0.2*\L) node[left, rotate=-45] (scrip) {\small{infalling null rays}};
   \draw[blue] (1.6*\L,-0.3*\L) node[left, rotate=+45] (scrip) {\small{$N_{ij}$}};
   %\draw[green] (0.2*\L,0.2*\L) node[below, rotate=+45] (scrip) {\small{$L_{ij}$}};
    \draw[green,->] (0.15*\L,0.1*\L) -- (0.35*\L,0.3*\L);
    \draw[blue] (-0.8*\L,-0.6*\L) node[left, rotate=45] (scrip) {{$r=0$}}; 
    \filldraw[black] (-0.42*\L,-0.42*\L) circle (2pt); \draw[magenta] (-0.4*\L,-0.4*\L) node[above, rotate=45] (scrip) {\small{${\cal N}_v$}}; 
    \draw[red] (0.9*\L,1*\L) node[left, rotate=45] (scrip) {{${\cal N}$}};
    \draw[brown] (0.5*\L,-0.5*\L) node[left] (scrip) {{$r>0$}};
    \draw[brown] (-0.2*\L,0.3*\L) node[left] (scrip) {{$r<0$}};
\end{tikzpicture}
\caption{${\cal N}$ is a null boundary  at $r=0$. $v$ is the null coordinate along ${\cal N}$ and the $D-2$ dimensional ``transverse'' space ${\cal N}_v$, constant $v$ surfaces on ${\cal N}$,  is spanned by coordinates $x^i, i=1,2,\cdots, D-2$. The null boundary ${\cal N}$ does not necessarily have an initial or endpoint. We  excise the $r<0$ region and formulate physics in $r\geq 0$. $N_{ij}$, also called (Bondi) news, parameterize infalling null rays. 
%while $L_{ij}$ are null rays propagating along the boundary.  
The passage of $N_{ij}$ through ${\cal N}$ is interpreted as dissolution of gravitons onto the boundary from the viewpoint of observers in $r\geq 0$ region. 
}\label{Fig:null boundary}
\end{figure}

\paragraph{Null boundary.} Among different choices for the boundary, we consider a null boundary ${\cal N}$, which we take to be $r=0$ surface, cf. Fig. \ref{Fig:null boundary}. Any accelerated observer finds such a null boundary, its Rindler horizon. This choice is  also motivated by the questions regarding black holes, where the boundary models the black hole horizon.  The null boundary is special as it only allows for a one-way passage of the null rays to the behind ($r<0$) region. 

${\cal N}$ is a null surface which  is topologically $\mathbb{R}_v\ltimes {\cal N}_v$. In what follows we view $v$ as the ``time'' coordinate for the boundary observers, ${\cal D}_v$ denotes the covariant time derivative along ${\cal N}$ and $x^i$ span ${\cal N}_v$. Being a null surface, the metric on ${\cal N}$ is degenerate and ${\cal N}$ may be specified by the metric on ${\cal N}_v$ $\Omega_{ij}$ and a vector $l^\mu$ which is null. Moreover, we also need to define the covariant derivatives on ${\cal N}$;  we denote covariant derivative along the null direction $v$ by ${\cal D}_v$ and covariant derivative along $x^i$ directions by $\nabla_i$. We choose $\nabla_i$ to be compatible with the metric on ${\cal N}_v$, $\Omega_{ij}$.  
${\cal N}$ can be locally obtained as speed of light to zero limit of a $D-1$ Minkowski space, i.e. a Carrollian spacetime \cite{Levy-Leblond, SenGupta, Levy-Leblond-2, Duval:2014uoa, Duval:2014uva, Morand:2018tke, Ciambelli:2019lap}.

Any two points  $(v_1,x^i_1), (v_2, x_2^i)$ on ${\cal N}$ are out of relativistic causal contact, unless $x^i_1=x^i_2$. So,  information on these points can't be connected by a causal dynamics and  the theory on ${\cal N}$ does not have  a relativistic description; it is a Carrollian local field theory, see \cite{Duval:2014lpa, Duval:2017els, Ciambelli:2018wre, Ciambelli:2018xat, Ciambelli:2018ojf, Bagchi:2019xfx, Donnay:2019jiz} and references therein.

\paragraph{Null boundary symmetries.} Choosing the null boundary ${\cal N}$ as described above, partially fixes $D$ dimensional diffeomorphisms to $D-1$ diffeomorphisms on ${\cal N}$ plus  local scaling of the $r$ coordinate, $r\to W(v,x^i)r$. Explicitly, the symmetry generators are specified by 
\begin{equation}\label{diffeos}
\begin{split}
    v &\to v+ T(v,x^i),\\
    r &\to W(v,x^i) r,\\
    x^i &\to x^i+ Y^i(v,x^j).
\end{split}    
\end{equation}
The above are $D-1$ foliation preserving diffeomorphisms,  local translations in $v$ and $x^i$, plus $W(v,x^i)$. Since $\partial_r$ is a null direction, $W$ generates local boosts on ${\cal N}$. Boosts along $x^i$ directions do not keep ${\cal N}$ null and are not among our symmetry generators. Therefore, the boundary theory is expected to have ``$D-1$ dimensional conformal Carrollian'' symmetry \cite{Duval:2014lpa} as a local symmetry. Here, we focus on the physical picture emerging from recent papers \cite{Donnay:2019jiz,  Adami:2020ugu, Adami:2021sko} and in particular \cite{Adami:2021nnf, Adami:2021kvx}, without delving into interesting technicalities of  the analyses.  For a detailed analysis  one may look at those papers.  For analysis of boundary symmetries and charges for null boundaries, see also \cite{Donnay:2015abr, Chandrasekaran:2018aop, Chandrasekaran:2019ewn, Grumiller:2019fmp, Adami:2020amw}.

\paragraph{Boundary degrees of freedom.} Front observers, observers in $r>0$ region,  may see things falling in, but not coming out. We are going to excise $r<0$ region and only focus on $r\geq 0$ region. Front observers interpret an infalling flux  as something ``dissolving'' into the null boundary. In order this picture to physically make sense one should add appropriate boundary degrees of freedom (b.d.o.f) which reside on ${\cal N}$. Their role is to compensate for the d.o.f. in behind the boundary $r<0$ region which has been excised from the spacetime, such that the front observer can provide a unitary description. That such a description  exists, is implied by the Einstein's equivalence principle and that different observers should  have access to the complete description of the events in their causally accessible region. (Recall that as discussed above,  ${\cal N}$ can be viewed as Rindler horizon of a congruence of locally accelerated observers.)

To perform the role they are supposed to, b.d.o.f should readjust themselves as a response to the dissolution of the flux onto the boundary. This readjustment is governed by the  flux-balance equations which are simply (Einstein) field equations projected along and computed at ${\cal N}$, the Raychaudhuri and Damour equations at ${\cal N}$. There are $D-1$ such equations \cite{Adami:2021nnf, Adami:2021kvx}.  One should note that while the details of these equations do depend on the gravity theory we are considering, their existence and that they are just first order differential equations in time $v$, is merely a consequence of diffeomorphism invariance of the theory and do not depend on the theory. 
%Being   first order equations, the flux-balance equations may also be interpreted as diffusion equation from boundary viewpoint.

\paragraph{Null boundary solution space.} One may construct space of all solutions to Einstein GR with ${\cal N}$ as the null boundary through a perturbative expansion in $r$ \cite{Adami:2021nnf}. This solution space is specified by $D$ functions over ${\cal N}$, namely $D$ arbitrary functions of $v$ and transverse coordinates $x^i$, plus the bulk graviton modes which can propagate in the bulk. 
%In what follows we denote the metric on codimension 2 surface ${\cal N}_v$ by $\Omega_{ij}(v,x^i)$ and $\Omega(v,x^i)=\sqrt{\det{\Omega_{ij}}}$. 
\begin{itemize}
    \item \textbf{Boundary modes.} The $D$ b.d.o.f may be labeled by the set of $D$ charges ${\cal Q}_A(v,x^i), A=1,2,\cdots, D$, associated with and in one-to-one correspondence to, the $D$ residual diffeomorphisms in \eqref{diffeos}. ${\cal Q}_A$ consist of two ``scalar'' modes $\Omega(v,x^i), {\cal P}(v,x^i)$ and a ``vector'' mode ${\cal J}_i(v,x^i)$. $\Omega(v,x^i):=\sqrt{\det{\Omega_{ij}}}$ is the charge associated with the local boosts at ${\cal N}$ (local Carrollian scaling) $W(v,x^i)$, ${\cal P}$ associated with ``supertranslations'' along $v$,   and ${\cal J}_i(v,x^i)$  with the ``superrotations'' $Y^i(v,x^j)$.
\item \textbf{Bulk modes.} The graviton modes fall into two classes, parametrized by symmetric traceless tensors $N_{ij}=N_{ij}(v,x^i), L_{ij}=L_{ij}(r, x^i)$ \cite{Adami:2021nnf}, {cf.} Fig. \ref{Fig:null boundary}. $N_{ij}$ is the trace-free part of ${\cal D}_v \Omega_{ij}$ and  $L_{ij}$ modes vanishes at $r=0$ boundary and are $v$ independent, therefore $L_{ij}$ are not a part of boundary data and do not affect boundary dynamics. 
\end{itemize}
The flux-balance equations involve  first order $v$ derivatives of the boundary modes and  $N_{ij}$, and not $L_{ij}$.  These equations from the boundary observer viewpoint are  like an ordinary diffusion equation, describing how the news $N_{ij}$ dissolves/diffuses as it reaches the boundary. The same equations can be interpreted  as ``null boundary memory effect'' as they tell us how the news $N_{ij}$ is encoded into the b.d.o.f after its dissolution. The boundary memory  is a local effect on ${\cal N}_v$, while it involves an integration over $v$. See \cite{Adami:2021nnf} for the details of  analysis.

\paragraph{Null boundary phase space.} Solution space is a phase space equipped with a symplectic two-form $\boldsymbol{\Omega}$. One may workout this symplectic form using covariant phase space formalism, see \cite{The-BH-book} for a detailed review. The result of the analysis is \cite{Adami:2021kvx} 
\be\label{sympl-two-form} 
\boldsymbol{\Omega}=\frac{1}{16\pi G} \int_{\cal N} \ \sum_{A=1}^D\ \delta {\cal Q}_A\wedge \delta \mu^A +  \delta (\Omega N_{ij})\wedge \delta \Omega^{ij}.
\ee
where $G$ is the Newton constant, $\delta X, \forall X$ is a one-form over the solution space and $\Omega^{ij}$ is inverse of metric $\Omega_{ij}$. $\mu^A=\mu^A(v,x^i)$ are canonical conjugates to the charges ${\cal Q}_A$ and are related to ${\cal Q}_A$ and the graviton modes $N_{ij}$ through the balance equations.\footnote{The flux-balance equations are algebraic equations for the canonical conjugates to $\Omega, {\cal J}_i$ charges.} The canonical conjugate to ${\cal P}$ is ${\cal D}_v\Omega=\Omega (v,x^i)\Theta (v,x^i)$ where $\Theta$ is the expansion of the null surface, the canonical conjugate to $\Omega$ is local acceleration of null rays generating ${\cal N}$, which we will denote by $\Gamma (v,x^i)$ and canonical conjugate to ${\cal J}_i$ are angular velocity of the same null rays which we will denote by ${\cal U}^i$. Therefore, in this conventions for any X, ${\cal D}_v X= \partial_v X+{\cal L}_{{\cal U}} X$, where ${\cal L}_{{\cal U}}$ denotes Lie derivative along ${\cal U}^i$. 

\paragraph{Boundary symplectic form.} As discussed the boundary and bulk modes are distinct, as the former are in scalar and vector representation of the ${\cal N}_v$ diffeomorphisms, while the latter is in (symmetric-traceless) tensor mode. The distinction between the two and the name boundary and bulk, can be made more explicit. Let us turn off the bulk modes and set $N_{ij}=0$. For this case, Raychaudhuri and Damour equations simplify to \cite{Adami:2021kvx}
\begin{equation}\label{EoM}
    {\cal D}_v {\cal P}= \Gamma (v,x^i), \qquad  {\cal D}_v\boldsymbol{{\cal J}}_i=0, \quad \boldsymbol{{\cal J}}_i:={\cal J}_i+ \nabla_i (\Omega {\cal P}),
\end{equation}
where $\nabla_i$ is the covariant derivative on ${\cal N}_v$ compatible with metric $\Omega_{ij}$. We stress that while we have taken $N_{ij}=0$, the expansion $\Theta={\cal D}_v \Omega/\Omega$ is taken to be non-zero.

The symplectic form \eqref{sympl-two-form} for this case takes the form
\be\label{sympl-two-form-b'dry-1} 
\begin{split}
\boldsymbol{\Omega}_{N_{ij}=0}
&=\frac{1}{16\pi G} \int_{\cal N} \  \left[ \delta ({\cal D}_v\Omega)\wedge \delta{\cal P} + \delta \Gamma \wedge \delta \Omega+ \delta {\cal U}^i\wedge \delta {\cal J}_i \right]
%\\
%&=\frac{1}{16\pi G} \int_{\cal N} \  {\cal D}_v\left( \delta\Omega \wedge \delta{\cal P} \right)+ \delta {\cal U}^i\wedge \delta\boldsymbol{{\cal J}}_i
\\ 
&=\frac{1}{16\pi G} \int_{{\cal N}}\ {\cal D}_v\left[\delta\Omega \wedge \delta{\cal P} + \delta \boldsymbol{\omega}^i\wedge \delta \boldsymbol{{\cal J}}_i \right]
\end{split}
\ee
where  we integrated by part, used \eqref{EoM} and  $\boldsymbol{\omega}^i=\int_\gamma \text{d}v\ {\cal U}^i$ where $\gamma (v)$ is an arbitrary path such that ${\cal D}_v\boldsymbol{\omega}^i= {\cal U}^i$. 
Based on the above, we can define the boundary symplectic form, 
\be\label{sympl-two-form-b'dry} \boxed{
\boldsymbol{\Omega}_{\text{b'dry}}=\frac{1}{16\pi G} \int_{{\cal N}_v}\ \left[\delta\Omega \wedge \delta{\cal P} + \delta \boldsymbol{\omega}^i\wedge \delta \boldsymbol{{\cal J}}_i \right].
}\ee
The important point in \eqref{sympl-two-form-b'dry} is that the symplectic form takes the form of a codimension 2 integral, an integral over the constant time $v$ slice, ${\cal N}_v$. That is, in $N_{ij}=0$ sector, $\boldsymbol{\Omega}_{\text{b'dry}}$ may be viewed as the symplectic form of a boundary theory which resides on ${\cal N}$. 

The above analysis has several interesting implications, some of which we discuss here. 
\begin{enumerate}
    \item In the absence of external flux,
$N_{ij}=0$, the boundary system is a closed system with a conserved symplectic form $\boldsymbol{\Omega}_{\text{b'dry}}$. 
\item There are $D-1$ b.d.o.f $\Omega, \boldsymbol{\omega}^i$  and their canonical conjugates are ${\cal P}, \boldsymbol{{\cal J}}_i$. 
\item The canonical equal time Poisson brackets of the system are
\be\label{Heisenberg}
\{\Omega(v,x^i), \Omega(v, y^i)\}=0=\{{\cal P}(v,x^i), {\cal P}(v, y^i)\},\qquad 
\{\Omega(v,x^i), {\cal P}(v, y^i)\}=\frac{1}{4G} \delta^{D-2}(x-y),
\ee
\be\label{J-omega}
\{ \boldsymbol{\omega}^i(v,x^i),  \boldsymbol{{\cal J}}_j(v, y^i)\}=\frac{1}{4G} \delta^i{}_j\ \delta^{D-2}(x-y).
\ee
and a closer inspection reveals that the Poisson bracket $\{{\cal J}_i(v,x^i), {\cal J}_j(v,y^i)\}$  takes the form of the algebra of $D-2$ dimensional diffeomorphisms for any $v$ \cite{Adami:2021kvx, Adami:2021nnf}. To see this we note that \eqref{EoM} implies, $\boldsymbol{\omega}^i=\boldsymbol{\omega}^i(\Omega, {\cal P}, \boldsymbol{{\cal J}}_j)$. 
\item That these Poisson brackets have the same form for any given $v$ is a manifestation of the fact the b.d.o.f can be defined at any given $v$, on the codimension two surface ${\cal N}_v$; explicitly, the d.o.f of the boundary theory are defined on corners, resonating the viewpoint advocated in some recent papers \cite{Ciambelli:2021nmv,Freidel:2021cjp}.  
\item We restress again that while $N_{ij}=0$ in this sector, $\Theta\neq0$. The vanishing expansion  $\Theta=0$ implies $N_{ij}=0$ \cite{Adami:2021nnf}, but the converse is not true.  For the $\Theta=0$ case, we are in a ``stationary phase space'' in which ${\cal P}$ is fixed and the phase space is reduced to the one specified by only $D-1$ charges, $\Omega(x^i), {\cal J}_j(x^i)$.
\item As argued, the  b.d.o.f can be governed by a well defined dynamics in $v$ which cannot be a relativistic one, it should be a Carrollian  evolution.
\item Our analysis specifies the phase space and symplectic form of the boundary theory but the dynamics (Hamiltonian) of this system is not specified through the boundary symmetry analysis we reviewed here. At this level it is free to be chosen; it may be fixed through some other physical requirements/criteria.
\item When $N_{ij}\neq 0$ the boundary theory is an open system due to the passage of the flux of gravitons through ${\cal N}$. In this case $\Gamma=\Gamma (\Omega,{\cal P}, {\cal J}_i; N_{ij})$ and $\boldsymbol{\omega}^i=\boldsymbol{\omega}^i(\Omega, {\cal P}, {{\cal J}}_j; N_{ij})$ and the symplectic form
\eqref{sympl-two-form} does not localize on a given $v$, it will have a boundary part (integral over ${\cal N}_v$) and a bulk part (integral over ${\cal N}$). Put differently, the boundary symplectic form $\boldsymbol{\Omega}_{\text{b'dry}}$ will not remain conserved and there is a symplectic flux proportional to $N_{ij}$.
\end{enumerate}

\paragraph{Null surface thermodynamics.} The above description of the solution space, especially noting that $\Omega$ is the charge associated to boosts on ${\cal N}$ and its canonical (thermodynamical) conjugate variable $\Gamma$ is local acceleration, suggests that there should be a thermodynamical interpretation. In this thermodynamical description, \textit{entropy density} at any constant $v$ on ${\cal N}$ is $4G \Omega$, extending seminal Wald's result \cite{Wald:1993nt, Iyer:1994ys}, and its conjugate variable $\Gamma$ is the \textit{local temperature} (times $4\pi$), extending seminal Unruh's analysis \cite{Unruh:1976db}. The other terms, too, have a natural thermodynamic description, with local first law, local Gibbs-Duhem and local zeroth law,  as established in \cite{Adami:2021kvx}. Here by local we mean local on ${\cal N}$. 

This is in general an open thermodynamical system as it can be out of (local) equilibrium due to the passage of news $N_{ij}$ or having a non-zero expansion $\Theta$; thermal equilibrium may be achieved only in the absence of news \cite{Adami:2021kvx}, when the boundary theory becomes a closed (isolated) thermodynamical system. We stress that balance equation which is  describing the rearrangement of b.d.o.f due to the passage of $N_{ij}$ through ${\cal N}$, should not be viewed as a (relativistic) dynamical equation. This rearrangement happens locally (instantaneously) at any given $v$  to ensure diffeomorphism invariance of the $D$ dimensional theory.

\paragraph{Concluding remarks.} To summarize, for any locally accelerated observer we need to formulate physics on one side of a null surface. This system is an open thermodynamic system; the dissolution of bulk infalling modes into this system is governed by the flux-balance equations. The configuration/phase space of the system is a direct sum of boundary and bulk modes. The boundary d.o.f may be parametrized by the area density $\Omega$ at a given $v$ and its canonical conjugate variable is ${\cal P}$, as is seen from \eqref{sympl-two-form-b'dry}.

This description is suggestive of a new ``semiclassical'' description of the system where the boundary mode is treated  quantum mechanically while the bulk mode $N_{ij}$ is kept classical. That is, we quantize the canonical Poisson brackets \eqref{Heisenberg} and \eqref{J-omega} by promoting these fields to operators and the Poisson brackets to commutators. In this system $1/(4G)$ effectively plays the role of  $\hbar$. As discussed the entropy density $S=\Omega/(4G)$. With the appropriate dynamics chosen for this boundary system $\Omega$ may be quantinzed in units of $\hbar$. This semiclassical description may be relevant to questions regarding black hole microstates and the information puzzle. We hope to report on this new semiclassical quantization in future publications. 

%\snote{word limit main text: 1500; circa 3 pages}

%\newpage

\paragraph{Dedication and connection to Paddy's works.} This work is dedicated to the memory of  T. Padmanabhan, Paddy, whose work has directly and indirectly influenced the current research  discussed here. Paddy had realized the crucial role of boundary conditions and boundary dynamics in his influential works on semiclassical and quantum aspects of black holes. In particular in \cite{Parattu:2015gga,Parattu:2016trq} the variational principle and the required boundary term  for the null surfaces was discussed and analyzed. These analysis was then used in his later works \cite{Chakraborty:2016dwb, Chakraborty:2019doh}, where it was argued that ``the null surfaces in spacetime exhibit (observer-dependent) thermodynamic features. This suggests a possible thermodynamic interpretation of the boundary term when the boundary is a null surface.'' These arguments resonates with analysis of \cite{Adami:2021kvx} and  our discussion above.

\paragraph{Acknowledgments.}

I thank  Alfredo Perez, S. Sadeghian, Ricardo Trancoso and especially Hamed Adami, Daniel Grumiller, Vahid Taghiloo, Hossein Yavartanoo and Celine Zwikel for collaboration on projects upon which this Letter is based. I would also like to thank Glenn Barnich, Laurent Freidel, Daniele Paranzetti, Romian Ruzziconi and  Ali Seraj for fruitful discussions. I would like to thank ICTP Trieste, ULB Bruxelles, TUW Vienna and organizers of IFPU (SISSA \& ICTP) workshop on ``Holography and gravitational waves'' where this work was presented and several instructive comments and discussions made. 
%DG was supported by the Austrian Science Fund (FWF), projects P 30822, P 32581 and P 33789.
The author acknowledges SarAmadan grant No. ISEF/M/401332.

The author is grateful to Sumanta Chakraborty, Dawood Kothawala, Sudipta Sarkar, Amitabh Virmani for their effort in putting together  the Topical Collection (TC) “In Memory of Prof. T. Padmanabhan” of the journal General Relativity and Gravitation (GERG).

%{\footnotesize

%\bibliographystyle{fullsort}
%\bibliography{reference}

\providecommand{\href}[2]{#2}\begingroup\raggedright\endgroup

%}

\end{document}